\RequirePackage{lineno}
\documentclass[twocolumn,aps,prl,groupedaddress,nofootinbib,showpacs]{revtex4-1}
\usepackage{graphicx}
\usepackage{amsmath}
\usepackage{bm}
\usepackage{slashed}
\usepackage{epsfig}
\usepackage{amsfonts}

\newcommand\gev{\mathrm{~GeV}}
\newcommand\tev{\mathrm{~TeV}}

\newcommand{\jpsi}{{J/\psi}}
\newcommand\psip{{\psi^\prime}}

\newcommand{\state}[4]{{^#1\hspace{-0.6mm}#2_{#3}^{[#4]}}}

\newcommand\CScSa{\state{3}{S}{1}{1}}

\newcommand\COaSz{\state{1}{S}{0}{8}}

\newcommand\COcSa{\state{3}{S}{1}{8}}
\newcommand\COcPz{\state{3}{P}{0}{8}}

\newcommand\COcPj{\state{3}{P}{J}{8}}

\newcommand\mo{{\mathcal O}}

\newcommand{\LDME}[2]{\langle\mo^{#1}(#2)\rangle}
\newcommand\mops{\LDME{\jpsi}{\CScSa}}
\newcommand\mopa{\LDME{\jpsi}{\COaSz}}
\newcommand\mopb{\LDME{\jpsi}{\COcSa}}
\newcommand\mopc{\LDME{\jpsi}{\COcPz}}

\newcommand\moppc{\LDME{\psip}{\COcPz}}

\newcommand{\vt}[1]{{{\boldsymbol #1}_\perp}}
\newcommand{\vtn}[2]{{{\boldsymbol #1}_{#2\perp}}}

\newcommand{\vx}{{\vt{x}}}
\newcommand{\vy}{{\vt{y}}}

\newcommand{\vvr}{{\vt{r}}}

\newcommand{\vl}{{\vt{l}}}
\newcommand{\vp}{{\vt{p}}}
\newcommand{\vk}{{\vt{k}}}
\newcommand{\vka}{{\vtn{k}{1}}}

\newcommand{\vtp}[1]{{{\boldsymbol #1}'_\perp}}

\newcommand{\vxp}{{\vtp{x}}}
\newcommand{\vyp}{{\vtp{y}}}

\newcommand{\vlp}{{\vtp{l}}}

\newcommand{\vkp}{{\vtp{k}}}

\begin{document}
\title{Comprehensive description of  $J/\psi$ production in proton-proton collisions at collider energies}

\author{Yan-Qing Ma and Raju Venugopalan}


\affiliation{Physics Department,
                Brookhaven National Laboratory,
                Upton, NY 11973-5000, USA}

\date{\today}

\begin{abstract}
We employ a small $x$ Color Glass Condensate (CGC)+ Non-Relativistic QCD (NRQCD) formalism to compute $J/\psi$ production at low $p_\perp$ in proton-proton collisions at collider energies. Very good agreement is obtained for total cross-sections, rapidity distributions and low momentum $p_\perp$ distributions. Similar agreement is obtained for $\psi^\prime$ production. We observe an overlap region in $p_\perp$ where our results match smoothly to those obtained in a next-to-leading order (NLO) collinearly factorized NRQCD formalism. The relative contribution of color singlet and color octet contributions can be quantified in the CGC+NRQCD framework, with the former contributing approximately $10\%$ of the total cross-section.
\end{abstract}

\maketitle


\allowdisplaybreaks

The production of heavy quarkonium states is
an excellent laboratory to understand hadronization in QCD~\cite{Brambilla:2010cs}. A significant development was the non-relativistic QCD (NRQCD) factorization formalism~\cite{Bodwin:1994jh} that provides systematic estimates of the magnitudes of universal long distance matrix elements (LDMEs) contributing to cross-sections of heavy quarkonium states. Based on this NRQCD framework, next-to-leading order (NLO) computations performed in recent years~\cite{Butenschoen:2012px,Chao:2012iv,Gong:2012ug} describe nearly all the high $p_\perp$ ($p_\perp >> M$, where $M$ is the quarkonium mass) data on heavy quarkonium production at hadron colliders, with further improvements anticipated from high $p_\perp$ logarithmic resummations~\cite{Kang:2014tta,Bodwin:2014gia,Ma:2014svb}

In contrast, heavy quarkonium production in the low $p_\perp \lesssim M$ region is far from understood. This regime dominates the total cross-section for production of heavy quarkonia at colliders.
For collider center of mass (c.m) energies $\sqrt{s}$, the dynamics is sensitive to large logarithms in $x\sim M/\sqrt{s}$. Summing these small $x$ logs leads to the phenomenon of gluon saturation,  characterized by a dynamically generated semi-hard scale $Q_S$ in the hadron wavefunctions~\cite{Gribov:1984tu,Mueller:1985wy}.
Recently in \cite{Kang:2013hta},
the Color Class Condensate (CGC) effective theory of small $x$ QCD ~\cite{Gelis:2010nm} results for the short distance heavy quark pair production cross-section~\cite{Blaizot:2004wv,Fujii:2005vj,Fujii:2006ab} were combined with the                                                                                                                                                                                                                                                                                                LDMEs of NRQCD to provide analytic expressions for a large number of quarkonium final states.

In this letter, we will provide first quantitative results in this novel CGC+NRQCD framework for $J/\psi$ production in proton-proton (p+p) collisions at collider energies\footnote{Previous CGC comparisons to quarkonia for p+p collisions employed the color evaporation model~\cite{Fujii:2013gxa}. For p+A collisions, see also \cite{Kharzeev:2012py}.}. We show that results in this framework can be matched smoothly to NLO perturbative QCD results at high $p_\perp$, thereby providing the missing link for a unified description for quarkonium production in all phase space. This unified framework is identical for the description of both p+p and proton-nucleus (p+A)  collisions. With some numerical effort, it can be extended systematically~\cite{Gelis:2005pb} to quarkonium production in the early stages of nucleus-nucleus (A+A) collisions, where quarkonium dissociation is a signature of quark-gluon plasma formation. A further nice feature of this framework is that allows one to quantify the relative contribution of color singlet and color octet channels to quarkonium final states in both p+p and p+A collisions.

In the NRQCD factorization formalism \cite{Bodwin:1994jh}, the inclusive production of a heavy quarkonium state $H$
is expressed as
\begin{align}\label{eq:factorization}
d\sigma_H = \sum_\kappa d\hat{\sigma}^{\kappa}\langle {\cal O}_\kappa^H \rangle,
\end{align}
where $\kappa=\state{{2S+1}}{L}{J}{c}$ are the quantum numbers of the produced intermediate heavy quark pair with standard spectroscopic notation for the angular momentum, and the superscript $c$ denotes the color state of the pair, which can be either color singlet (CS) with $c=1$ or color octet (CO) with $c=8$. For $\jpsi$ production, the primary focus here, the most important intermediate states are $\CScSa$, $\COaSz$, $\COcSa$ and $\COcPj$. In Eq. \eqref{eq:factorization}, $\langle {\cal O}_\kappa^H \rangle$ are non-perturbative but universal NRQCD LDMEs, which
can be extracted from data. The $d\hat{\sigma}^{\kappa}$ denote the short distance coefficients for the production of a heavy quark pair with quantum number $\kappa$. Based on the heavy quark pair production amplitude in \cite{Blaizot:2004wv,Fujii:2006ab}, $d\hat{\sigma}^{\kappa}$ have been calculated for all $S$-wave and $P$-wave intermediate states in \cite{Kang:2013hta}.

In the following, we will give general expressions for ``dilute-dense scattering". In the CGC power counting~\cite{Gelis:2010nm},   this could represent either p+A collisions or forward p+p collisions.
At collider energies,  $x$ in the forward going ``dilute" proton may be small enough that small $x$ resummation is relevant, but phase space densities are still small. Conversely, phase space densities in the ``dense" backward going proton can reach their maximal value of $1/\alpha_s$, indicating gluon saturation. Because large $x$ modes are probed at high $p_\perp \gg Q_S$, there is a limited range in $p_\perp$ where the CGC effective theory is valid. At such large $p_\perp$, as we shall discuss further, perturbative computations in the collinear factorization framework should be more reliable.

In this dilute dense CGC  framework, we can express $d\hat{\sigma}^{\kappa}$ in the color octet channel as \cite{Kang:2013hta}
\begin{align}\label{eq:dsktCO}
\begin{split}
\frac{d \hat{\sigma}^\kappa}{d^2\vp d
y}\overset{\text{CO}}=&\frac{\alpha_s (\pi R_p^2)}{(2\pi)^{7}
(N_c^2-1)} \underset{\vka,\vk}{\int}
\frac{\varphi_{p,y_p}(\vka)}{k_{1\perp}^2}\\
&\times \mathcal{N}_Y(\vk)\mathcal{N}_Y(\vp-\vka-\vk)
\,\Gamma^\kappa_8,
\end{split}
\end{align}
when $\kappa$ is CO, $\int_\vk\equiv\int {d^2\vk}$, $\vp$ ($y$) are transverse momentum (rapidity) of produced heavy quarkonium, $y_p\equiv\ln(1/x_p)$ [$Y\equiv\ln(1/x_A)$] is the rapidity of gluons coming from dilute proton [dense proton]. The expression for $\Gamma^\kappa_8$ can be found in \cite{Kang:2013hta}. Further,
\begin{align}\label{eq:nf}
\mathcal{N}_{Y}(\vk)=\mathcal{N}_{Y}(-\vk)\equiv \underset{\vvr}{\int} e^{i\vk\cdot\vvr} D_{Y, \vvr} \,,
\end{align}
where $D_{Y,\vvr} = \left\langle {\rm Tr}\left[V_F(0) V_F^\dagger(\vvr)\right]\right\rangle_Y/N_c$ is the dipole forward scattering amplitude,  $V_F$ are lightlike Wilson lines in the fundamental representation, and $N_c$ the number of colors. Here $\langle\cdots\rangle_Y$ represents the renormalization group evolved  expectation value of this correlator in the target background field evaluated at rapidity $Y$. For more details, we refer the reader to \cite{Gelis:2010nm}. The unintegrated gluon distribution inside the proton is then expressed as
\begin{align}\label{eq:unintegrated}
\varphi_{p,y_p}(\vka)=\pi R_p^2 \frac{N_c k_{1\perp}^2}{4\alpha_s} \widetilde{\mathcal{N}}^A_{y_p}(\vka)\,,
\end{align}
where $\pi R_p^2$
is the effective transverse area of the proton, and $\widetilde{\mathcal{N}}^A$ is similar to ${\cal N}$ in Eq.~\eqref{eq:nf} but with its Wilson lines in the adjoint representation.

Interestingly, if $\kappa$ is CS, $d\hat{\sigma}^{\kappa}$ depends on both dipole correlators and novel quadrupole correlators \cite{Kang:2013hta}.
These latter are defined as $Q_{\vx \vxp \vyp \vy} = \left\langle {\rm Tr}\left[V_F(\vx) V_F^\dagger(\vxp) V_F(\vyp) V_F(\vy)\right]\right\rangle_Y/N_c\,$. Like the dipole correlators, they are universal gauge invariant quantities, and appear in a number of final states~\cite{JalilianMarian:2004da, Dominguez:2011wm}. The energy evolution of both dipole and quadrupole correlators is described by the Balitsky-JIMWLK hierarchy of small $x$ RG equations~\cite{Balitsky:1995ub,JalilianMarian:1997dw,Iancu:2000hn}. For the dipole correlator, in the large $N_c$ limit, the hierarchy has a closed form expression, the well known  Balitsky-Kovchegov (BK) equation~\cite{Balitsky:1995ub,Kovchegov:1999yj}. This equation can be solved numerically and is widely used in phenomenological applications.

In contrast, no such simple form exists
for the quadrupole correlator and solving the corresponding Balitsky-JIMWLK equation
is cumbersome for phenomenological applications. While the expressions simplify in a quasi-classical approximation  in the large $N_c$ limit~\cite{Blaizot:2004wv,Dominguez:2011wm}, the result is still too complicated for our purposes. For our study here,  we discovered an approximate factorized expression for the quadrupole correlator,
\begin{align}\label{eq:quadrupole}
&\hspace{-0.3cm}Q_{\vx \vxp \vyp \vy}\approx  D_{\vx-\vxp}D_{\vyp-\vy}\nonumber\\
&-D_{\vx-\vyp}D_{\vxp-\vy}+D_{\vx-\vy}D_{\vxp-\vyp}\nonumber\\
&+ \frac{1}{2} ( D_{\vx-\vyp}D_{\vxp-\vy} - D_{\vx-\vy}D_{\vxp-\vyp} )\nonumber\\
&\times (D_{\vxp-\vy}-D_{\vyp-\vy} + D_{\vyp-\vx}-D_{\vxp-\vx}) \,.
\end{align}
This result is exact when any two adjacent positions coincide: $\vx=\vxp$, $\vxp=\vyp$, $\vyp=\vy$ or $\vy=\vx$. We have checked~\cite{mv} that it is a good approximation to the Balitsky-JIMWLK results in \cite{Dumitru:2011vk}.

Making use of  Eq.~\eqref{eq:quadrupole}, the short distance coefficients for the CS channels in \cite{Kang:2013hta} simplify  significantly to
\begin{align}\label{eq:dsktCS}
\begin{split}
\frac{d \hat{\sigma}^\kappa}{d^2\vp d
y}\overset{\text{CS}}=&\frac{\alpha_s (\pi R_p^2)}{(2\pi)^{9}
(N_c^2-1)} \underset{\vka,\vk,\vkp}{\int}
\frac{\varphi_{p,y_p}(\vka)}{k_{1\perp}^2}\\
&\hspace{-1.5cm}\times \mathcal{N}_{Y}(\vk)\mathcal{N}_{Y}(\vkp)\mathcal{N}_{Y}(\vp-\vka-\vk-\vkp)\,
{\cal G}^\kappa_1,
\end{split}
\end{align}
where ${\cal G }$ has a different functional form from $\Gamma$ in \cite{Kang:2013hta} because of the above simplification. In particular, for the $\CScSa$ channel,
\begin{align}\label{eq:gmma3s11}
\begin{split}
{\cal G}^{\CScSa}_1=&\frac{k_{1\perp}^2\left( k_{1\perp}^2 + 4m^2\right)}{12 m}\\
&\hspace{-0.5cm}\times\left(\frac{1}{l_{\perp}^2+k_{1\perp}^2/4+m^2}-\frac{1}{{l'}_{\perp}^2+k_{1\perp}^2/4+m^2}\right)^2,
\end{split}
\end{align}
with $\vl=\vk-\frac{\vp-\vka}{2}, ~~ \vlp=\vkp-\frac{\vp-\vka}{2}$.

The key ingredient in both Eqs.~(\ref{eq:dsktCO}) and (\ref{eq:dsktCS}) is $\mathcal{N}_{Y}(\vk)$, which is obtained by solving the running coupling BK (rcBK) equation directly in momentum space. We use McLerran-Venugopalan (MV) initial conditions~\cite{McLerran:1993ni,McLerran:1993ka} for the dipole amplitude at the initial rapidity scale $Y_0\equiv\ln(1/x_0)$ (with $x_0=0.01$) for small $x$ evolution. Since the structure function $F_2$ in deeply inelastic scattering (DIS) is directly propotional (at small $x$) to the dipole amplitude, all the parameters in the rcBK equation are fixed from fits to the HERA DIS data~\cite{Albacete:2012xq}.
Other initial conditions discussed in the literature \cite{Albacete:2014fwa} give similar results at low $p_\perp$ for heavy quark pair production but differ significantly at high $p_\perp$ where the present formalism is not reliable no matter which initial condition is used. While the formalism captures the dominant higher order corrections at small $x$~\cite{Catani:1990xk,Catani:1990eg}, this is less assured as one goes to larger $x$ with increasing $p_\perp$.
Because NLO CGC expressions for Eqs.~\eqref{eq:dsktCS} and \eqref{eq:dsktCO} that extend their $p_\perp$ range are not available, we choose to restrict data comparisons to the more reliable small to moderate $p_\perp$ region. We shall explore
whether there is a $p_\perp$ range where our results overlap with the successful high $p_\perp$ NLO
collinear factorization approach \cite{Ma:2010yw,Butenschoen:2010rq}.

A similar issue of matching to the collinear framework occurs when $x >x_0$ is accessed respectively in the forward or backward going protons. Since the rcBK equation is solved only for $x<x_0$, we devised a scheme to extrapolate the dipole amplitude to larger $x$, using results for the collinear factorized gluon distributions (pdfs). From the relation between the pdfs and the unintegrated gluon distribution \cite{Fujii:2006ab,Kang:2013hta}, using  Eq.~\eqref{eq:unintegrated}, one obtains
\begin{align}\label{eq:match}
x_p f_{p/g}(x_p,Q^2)\overset{x_p\leq x_0}{=} a(x_p) \,C \int^{Q^2}\hspace{-0.4cm}k_{1\perp}^2 \widetilde{\mathcal{N}}^A_{y_p}(\vka) dk_{1\perp}^2\,,
\end{align}
where $C\equiv\frac{\pi R_p^2}{4\pi^3} \frac{N_c }{4\alpha_s}$ and $a(x_p)$ is an extra free function that we will now determine. We assume $a(x_p)\approx1$ when $x_p$ is close to $x_0$, such that  $a(x_0)=1$ and $a'(x_0)=0$. These two conditions result in two equations, which allow us to simultaneously determine $R_p$ and $Q$. If we use the CTEQ6M pdfs~\cite{Pumplin:2002vw} for $f_{p/g}(x_p,Q^2)$ and the rcBK equation with the MV initial condition to calculate $\widetilde{\mathcal{N}}^A_{y_p}(\vka)$, we determine $Q=Q_0=5.1\gev$ and $R_p=0.48$ fm. (Interestingly, this value for  $R_p$ is in very good agreement with the transverse gluon radius of the proton extracted from HERA diffractive data~\cite{Caldwell:2010zza}.)
Having fixed $R_p$ and $Q_0$ with our matching conditions, we use the extrapolation
\begin{align}\label{eq:extra}
\widetilde{\mathcal{N}}^A_{y_p}(\vka)\overset{x_p> x_0}{=} a(x_p) \widetilde{\mathcal{N}}^A_{Y_0}(\vka)\,,
\end{align}
with $a(x_p)$ determined to be
\begin{align}\label{eq:match2}
a(x_p)= x_p f_{p/g}(x_p,Q_0^2) \Big[C \int^{Q_0^2}\hspace{-0.4cm}k_{1\perp}^2 \widetilde{\mathcal{N}}^A_{Y_0}(\vka) dk_{1\perp}^2\Big]^{-1}\hspace{-0.2cm}.
\end{align}

The dilute-dense approximation employed here will break down when the phase space densities in both protons are large. This may be the case at central rapidities, especially at the LHC. In the CGC framework, quark pair production in this dense-dense regime cannot be computed analytically but may be feasible numerically. This is however beyond the scope of the present work and is a further source of systematic uncertainty.

Before we confront our framework to the data, we need to fix the charm quark mass and determine the LDMEs. We set the charm quark mass to be $m=1.5\gev$, approximately one half the $J/\psi$ mass.
The CO LDMEs were extracted in the NLO collinear factorized NRQCD formalism~\cite{Chao:2012iv} by fitting Tevatron high $p_\perp$ prompt $\jpsi$ production data; this gives $\mops=1.16/(2N_c) \gev^3$, $\mopa=0.089\pm0.0098 \gev^3$, $\mopb=0.0030\pm0.0012 \gev^3$ and $\mopc=0.0056\pm0.0021\gev^3$.
Note that the high sensitivity of short distance cross-sections to quark mass
is mitigated by the mass dependence of the LDMEs. The latter are extracted from particular data sets using Eq.~(\ref{eq:factorization}). For a different quark mass, the LDMEs extracted would be quite different~\cite{Ma:2010yw}, thereby ensuring only a weak quark mass uncertainty to NRQCD predictions.
We also note that while other NLO NRQCD fits exist, they are not suitable for our use.
For example, the fit in \cite{Bodwin:2014gia} relies on a cancellation between the $\COcSa$ channel and $\COcPj$ channels, which occurs only at NLO where short distance coefficients can be negative. To account for the uncertainties outlined, as well as higher order in $\alpha_s$ corrections, we introduce a systematic uncertainty of $30\%$ on top of the statistical uncertainties from the LDMEs.

\begin{figure}[!htbp]
\begin{center}
\includegraphics*[scale=0.75]{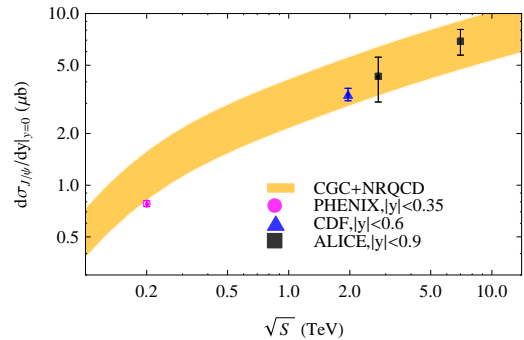}
\caption{The $\jpsi$ total production cross section at hadron colliders in the CGC+NRQCD framework compared to data at midrapidity. The data  are from \cite{Adare:2011vq,Acosta:2004yw,Abelev:2012kr,Aamodt:2011gj}.}
 \label{fig:SS}
 \end{center}
\end{figure}
\begin{figure}[!htbp]
\begin{center}
\includegraphics*[scale=0.75]{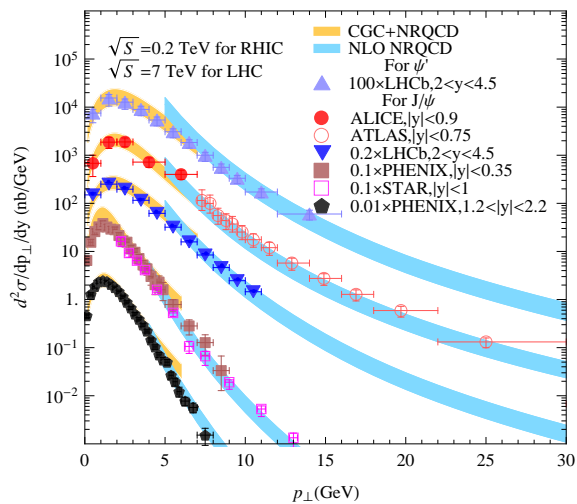}
\caption{The $\psip$ (top curve) and $\jpsi$ (other four curves) differential cross section as a function of $p_\perp$. Data from \cite{Adare:2011vq,Aamodt:2011gj,Aad:2011sp,Aaij:2011jh,Adamczyk:2012ey,Aaij:2012ag}. NLO NRQCD predictions are taken from~\cite{Ma:2010jj}. }
 \label{fig:pt}
 \end{center}
\end{figure}

In Fig.~\ref{fig:SS}, we compare our results for the $\jpsi$ total cross section at mid-rapidity for $\sqrt{S}$ from $0.2\tev$ to $7\tev$. In general, our predictions are consistent with the collider data. For the top RHIC energy, our result is a little larger then the PHENIX data.
However, (see Fig.~\ref{fig:pt}),  we are able to describe the PHENIX data on the $p_\perp$ dependence at forward rapidity.
The small discrepancy with PHENIX data at central rapidities may reflect the fact that our formalism is most reliable for large c.m energy or for forward rapidities.
As stated previously,
our computation is the first that includes contributions from both the color singlet and color octet channels to the total cross-section. In contrast to the Color Singlet Model
(CSM)~\cite{Chang:1979nn,Baier:1981uk} prediction, we find that in our formalism the CS contribution is  only  10\% of the total cross section. Our estimates show that for a large nucleus this relative contribution will rise to approximately 15\%- 20\%.

\begin{figure}[!htbp]
\begin{center}
\includegraphics*[scale=0.75]{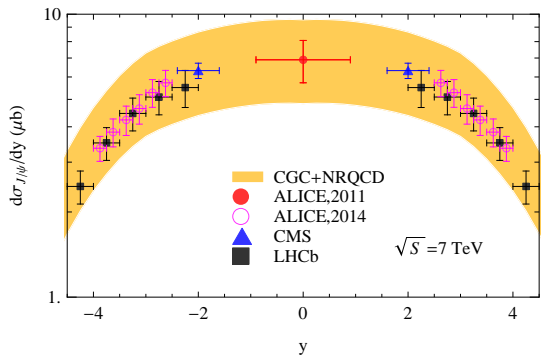}
\caption{$\jpsi$ differential cross section as a function of rapidity at LHC. Data from \cite{Aamodt:2011gj,Abelev:2014qha,Khachatryan:2010yr,Aaij:2011jh}.}
 \label{fig:y}
 \end{center}
\end{figure}
Our results for the differential cross section as a function of rapidity compared to $7\tev$ data at LHC are shown in Fig.~\ref{fig:y}.  All the  data lie well within the error band. At the larger rapidities, the results become sensitive to $x > 0.01$ in either the projectile or target (and conversely very small $x$ in
the target or projectile). At these forward (or backward) rapidities, as discussed previously, the matching of the dipole amplitude to the unintegrated pdfs at large $x$ is important. Note that this matching condition allowed us to fix the radius $R_p$ which was the only normalization parameter in Eqs.~(\ref{eq:dsktCO}) and (\ref{eq:dsktCS}) -- the apparent $\alpha_s$ dependence cancels out when Eq.~(\ref{eq:unintegrated}) is substituted in these equations. It is therefore striking that our results explain both the overall normalization at $Y=0$ and the relative normalization at forward rapidities after matching the pdfs smoothly to the dipole amplitude at $x=0.01$.

In Fig.~\ref{fig:pt}, we compare our results for $J/\psi$ differential cross section as a function of $p_\perp$ with experimental data at several c.m energies and rapidity regions. It is clear that small $p_\perp$ data is well described by our CGC+NRQCD formalism. Further, as anticipated in the previous discussion, the results in this formalism begin to disagree with data at higher $p_\perp$. From previous experience with single inclusive LHC data~\cite{Albacete:2012xq}, one expects to overshoot the data with MV initial conditions--this is precisely what we see. In Fig.~\ref{fig:pt}, we also show the NLO collinear factorized NRQCD results~\cite{Ma:2010jj} which show good agreement with data at large $p_\perp$. It is very interesting to observe an overlap region around $p_\perp\sim 5-6\gev$, which can be described by both the  CGC+NRQCD formalism and the NLO collinearly factorized NRQCD formalism. A good matching of the small $x$ and collinear factorized formalisms at large $p_\perp$ is seen in single inclusive hadron production by imposing exact kinematic constraints in the small $x$ formalism~\cite{Stasto:2014sea}. Since leading order collinear factorization results for heavy quark pair production are obtained as a limit of the CGC result~\cite{Gelis:2003vh}, imposing exact kinematics may help better understand the overlap between the two formalisms. In the low
$p_\perp$ region, a further refinement of our formalism will include resummation of logarithms of $p_\perp/M$ for $p_\perp \ll M$~\cite{Berger:2004cc,Sun:2012vc,Qiu:2013qka}.

Most of the experimental data presented are for inclusive $\jpsi$ production. These include $J/\psi$'s produced from $B$-meson decays as well as prompt production of $J/\psi$'s. The latter includes feeddown from higher excited charmonium states as well as direct $\jpsi$ production. However we only considered direct $\jpsi$ production contribution in our  theory results. Nevertheless, the comparison is meaningful. Firstly, the $B$-meson decay contribution in the small $p_\perp$ region is small, of order less than 10\%. Secondly, the LDMEs in \cite{Chao:2012iv} are obtained by fitting prompt $\jpsi$ data. Thus feeddown contributions are already roughly estimated in our results. With the expressions in \cite{Kang:2013hta} for the higher charmonium states, a fully consistent treatment of prompt $J/\psi$ production data is feasible in the near future. As a first step, we compare our results for the $\psi^\prime$ differential cross section as a function of $p_\perp$ with data in Fig.~\ref{fig:pt}.
In this comparison, we set the charm quark mass to be $m=M_{\psip}/2\approx1.84\gev$
and used the CO LDMEs extracted in~\cite{Ma:2010jj}. Theory and data agree well. However, if we set $m=1.5\gev$ for $\psi^\prime$, the results overshoot the data.  As noted, one anticipates the sensitivity to the quark mass in the short distance cross sections to be offset by their dependence in the LDMEs. However, for $\psi^\prime$, only two linear combinations of the three CO LDMEs of $\psip$ can be determined~\cite{Ma:2010jj}; because it is unconstrained, we  set $\moppc=0$ here. Thus $\psi^\prime$ has larger systematic uncertainties relative to $\jpsi$ that subsume the sensitivity of results to the quark mass. Based on our present work, all three $\psip$ CO LDMEs can be determined from a global fit. A consistent treatment of all charmonium and bottomonium states is in progress \cite{mv}. In the latter case,
we anticipate a larger contribution from the logarithmic resummation of \cite{Berger:2004cc,Sun:2012vc,Qiu:2013qka}.

Comparing our results to related work, $\jpsi$ total cross sections were studied recently in the CSM within the broad framework of collinear factorization~\cite{Brodsky:2009cf,Lansberg:2010cn}.  While the trend of the total cross section is described, the theory uncertainties have to be as large as a factor of 10 for data to lie within errors. Likewise, the CEM with collinear factorization~\cite{Nelson:2012bc} describes the total cross section, rapidity distributions and $p_\perp$ distributions with much larger uncertainties than our framework.

The results presented here are timely given the wealth of data extant and anticipated from RHIC to LHC for p+p and p+A collisions. When combined with developments in collinearly factorized NRQCD, they set the groundwork for a comprehensive understanding of the mechanism of heavy quark hadronization in QCD.

\section*{Acknowledgments}

We thank Adrian Dumitru, Kevin Dusling and Yasushi Nara for help with their rcBK codes. We thank Zhongbo Kang for his contributions at early stages of this work, and Jean-Philippe Lansberg and Lijuan Ruan for helpful discussions.  This work was supported in part by the U. S. Department of Energy under contract No. DE-AC02-98CH10886.

\providecommand{\href}[2]{#2}\begingroup\raggedright\endgroup

\end{document}